\documentclass[reprint,apl,aip,superscriptaddress,floatfix]{revtex4-1}
\usepackage{graphicx}

\usepackage{natbib}
\usepackage[usenames,dvipsnames]{pstricks}
\usepackage{pst-plot}
\usepackage{epsfig}
\usepackage{pst-grad} 
\usepackage{pst-plot} 
\usepackage{pstricks-add}
\usepackage{lmodern} 
\newcommand{\mycol}{1}
\newcommand{\beq}{\begin{equation}}
\newcommand{\eeq}{\end{equation}}
\usepackage{amsmath}
\usepackage{textgreek}

\begin{document}

\title{Sensitive spin detection using an on-chip SQUID-waveguide resonator}

\author{G. Yue}
\email{gy10c@my.fsu.edu}
\affiliation{Department of Physics, Florida State University, Tallahassee, Florida 32310, USA}
\affiliation{National High Magnetic Field Laboratory, Florida State University, Tallahassee, Florida 32310, USA}
\author{L. Chen}
\affiliation{CENSE, State Key Laboratory of Functional Material for Informatics, SIMIT, Chinese Academy of Sciences, Shanghai 200050, China}
\author{J. Barreda}
\affiliation{Department of Physics, Florida State University, Tallahassee, Florida 32310, USA}
\author{V. Bevara}
\affiliation{Department of Electrical and Computer Engineering, FAMU-FSU College of Engineering, Tallahassee, Florida 32310}
\affiliation{National High Magnetic Field Laboratory, Florida State University, Tallahassee, Florida 32310, USA}
\author{L. Hu}
\affiliation{Department of Physics, Florida State University, Tallahassee, Florida 32310, USA}
\author{L. Wu}
\affiliation{CENSE, State Key Laboratory of Functional Material for Informatics, SIMIT, Chinese Academy of Sciences, Shanghai 200050, China}
\author{Z. Wang}
\affiliation{CENSE, State Key Laboratory of Functional Material for Informatics, SIMIT, Chinese Academy of Sciences, Shanghai 200050, China}
\author{P. Andrei}
\affiliation{Department of Electrical and Computer Engineering, FAMU-FSU College of Engineering, Tallahassee, Florida 32310}
\affiliation{National High Magnetic Field Laboratory, Florida State University, Tallahassee, Florida 32310, USA}
\author{S. Bertaina}
\affiliation{Aix-Marseille Universit\'{e}, CNRS, IM2NP (UMR 7334), Marseille, France}
\author{I. Chiorescu}
\email{ic@magnet.fsu.edu}
\affiliation{Department of Physics, Florida State University, Tallahassee, Florida 32310, USA}
\affiliation{National High Magnetic Field Laboratory, Florida State University, Tallahassee, Florida 32310, USA}

\date{\today}

\begin{abstract} 
	Precise detection of spin resonance is of paramount importance to achieve coherent spin control in quantum computing. We present a novel setup for spin resonance measurements, which uses a dc-SQUID flux detector coupled to an antenna from a coplanar waveguide. The SQUID and the waveguide are fabricated from 20~nm Nb thin film, allowing high magnetic field operation  with the field applied parallel to the chip. We observe a resonance signal between the first and third excited states of Gd spins $S=7/2$ in a CaWO$_4$ crystal, relevant for state control in multi-level systems.  
\end{abstract}


\maketitle


Solid state spin-based qubits are studied for quantum computing due to their relatively long coherence time\cite{Morton2011,Gershenfeld1997}. Typical implementations of these qubits are molecule-based magnets\cite{Bertaina2008,Ardavan2007,Shiddiq2016,Bader2014}, nitrogen-vacancy (NV) centers in diamond\cite{Jelezko2004} and quantum spins in crystals\cite{Bertaina2007,Bertaina2009,Nellutla2007,Baibekov2017}.
These spin-based qubits are designed such that the spins are well separated in the crystal, leading to an increased decoherence time due to weak spin dipolar interactions.

Among the rare-earth ions, S-state lanthanide ions doped in a crystal have a rich energy level structure due to their large spin. Multi-level systems are promising for implementing few-qubits algorithms\cite{Leuenberger2001} or as quantum memories\cite{Blencowe2010,Chiorescu2010,Schuster2010,Wu2010}. For quantum technology applications, a higher sensitivity electron spin resonance (ESR) measurement is needed to be able to manipulate spins in mesoscopic crystals placed on superconducting chips\cite{Saito_2013,Bienfait2016,Abeywardana_2016,Antler_2013,Sigillito_2014}. 

\begin{figure}
	\centering
	\includegraphics[width=\mycol\columnwidth]{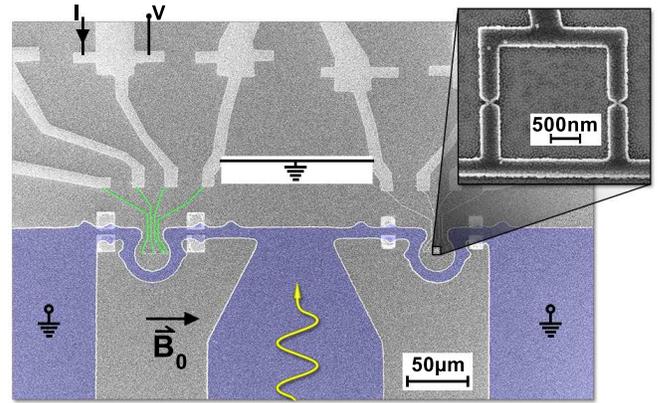}
	\caption{(color online) Scanning electron micrograph of the device and of the micro-SQUID (inset). The darker (blue) area is the coplanar waveguide with the two $\Omega$-loop shortcircuits between the central line and the lateral ground planes. The microwave excitation is depicted by the curvy arrow and an in-plane field $\vec{B_0}$ is generated by an external coil. The SQUIDs share a common ground (sketched as a white rectangle). Each SQUID has one $I-V$ line shown in green for clarity in their narrowest region.} \label{fig:device}
\end{figure}

Compared to other ultra-high sensitivity ESR measurements\cite{Baumann2015,Artzi2015}, the use of Josephson junctions can increase significantly the spatial resolution of the magnetic detection while allowing an on-chip implementation. For instance, the magnetic signal of one nanoparticle is detectable if placed on the junction of a micrometer sized superconducting quantum interference device (micro-SQUID)\cite{WW_PRL97}. We present a novel setup for ESR measurements, which combines the high spin sensitivity of an on-chip micro-SQUID and the flexibility of a coplanar waveguide for microwave excitation. Different dc-SQUID implementations were also used to detect molecular\cite{WW_V15,Cage2005} and diluted\cite{Toida2016} spins. In our case, the coupling of the two devices generates a cavity effect, which amplifies the microwave power seen by the spins.  When using micro-SQUIDs, the samples are positioned close to their loop for increased sensitivity since the device can work under in-plane magnetic fields in the range of $\sim$Tesla \cite{Wernsdorfer1999,Chen2010,Chen2016}. 

Using this setup, we successfully measured the resonance signal of Gd$^{3+}$ $S=7/2$ ions diluted in a CaWO$_4$ single crystal with a concentration of $0.05\%$. Moreover, the resonance is between the first and third excited states at $T=0.5$~K, demonstrating high sensitivity for applications in spin control in multi-level quantum system.

The device is etched from a 20~nm thin film of Nb to allow its operation in static magnetic fields parallel to the film. The Nb is sputtered on a Si chip coated with 300~nm SiO$_2$. A broadband 50~$\Omega$ coplanar waveguide is fabricated using ultra-violet lithography. The coplanar waveguide sends the microwave excitation into a central line of width 150~\textmugreek m, which is narrowed at the end and terminated with two shortcircuits shaped like letter $\Omega$, with internal radius of 15~\textmugreek m, towards the lateral ground planes as shown in Fig.~\ref{fig:device}. The micro-SQUIDs are fabricated in the middle of the $\Omega$-loops by means of electron beam lithography. The SQUID loop is 2.2~\textmugreek m $\times$ 2.2~\textmugreek m with two Dayem bridge Josephson junctions\cite{Anderson1964} of size 100~nm$\times$100~nm (see Fig.~\ref{fig:device} inset). The middle wires provide the ground for all SQUIDs (see the horizontal line in the inset). Each loop contains three SQUIDs and each one can be individually read using the current-voltage line indicated in Fig.~\ref{fig:device}.

With this device, microwave pulses can generate a B-field component perpendicular to the chip plane in the middle of the $\Omega$-loop. In absence of a SQUID detector in the $\Omega$-loop, the waveguide is a 50~$\Omega$ broadband device. By inserting a SQUID, an inductive coupling with the $\Omega$-loop allows the microwave energy to drain into the SQUID. This induces a cavity effect with a strong mode defined by the length of the patterned waveguide and with regions of maximum B-field at its ends. One can thus place a sample containing a small number of spins atop of the loop, to be excited by the microwave field. By using an external superconducting coil, a magnetic field $\vec{B_0}$ is applied $||$ to SQUID plane. Such a planar field provides the desired Zeeman splitting for the spins while leaves the SQUID almost unaffected.


The functioning principle of a dc-SQUID is based on the well-known modulation of its switching current as a function of flux penetrating its loop, which has a period of one flux quanta $\Phi_0=h/(2e)$ with $h$ Planck's constant and $e$ the electron charge. 

\begin{figure}
	\centering
	\includegraphics[width=\mycol\columnwidth]{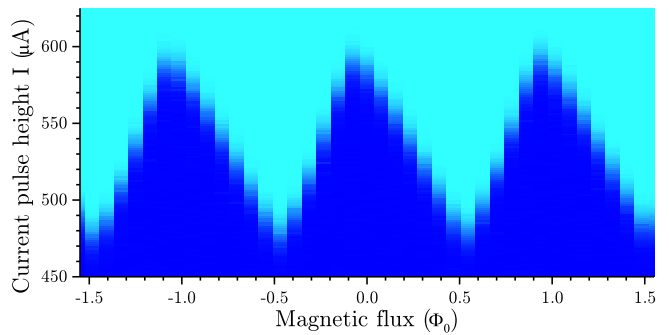}
	\caption{(color online) $P_{sw}$ at $T=4$~K as a function of bias current pulse and magnetic flux (dark blue for $P_{sw}=0$ and light blue for $P_{sw}=1$). The switching current, defined at $P_{sw}=0.5$, has a $\Phi_0$ periodicity with a depth of $\sim25\%$. }
	\label{fig:mod}
\end{figure}

The switching current is measured by monitoring the presence (switch event) or absence (non-switch event) of voltage pulses $V$ when a current pulse $I$ is injected in the SQUID (see the $I$ and $V$ lines in Fig.~\ref{fig:device}). The percentage of switching events defines the switching probability\cite{Chiorescu2003a} $P_{sw}$. A weak-link SQUID has a reduced depth of modulation, compared to a tunnel junction SQUID but can work under very large static fields. Fig. \ref{fig:mod} shows the $P_{sw}$ at $T=4$~K as a function of the current pulse height (vertical axis) and magnetic flux (horizontal axis). The switching current $I_{sw}$ is sharply defined as the pulse height such that $P_{sw}=50\%$, at the transition between the regions $P_{sw}=0$ (dark blue) and 1 (light blue). The modulation $I_{sw}(\Phi)$ allows the detection of a signal coming from a magnetic sample.

\begin{figure}
	\centering
	\includegraphics[width=\mycol\columnwidth]{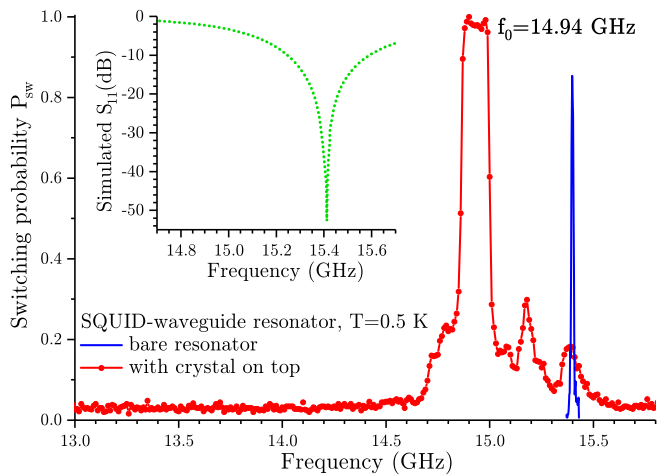}
	\caption{(color online) $P_{sw}$ of the SQUID at $T=$0.5~K under microwave excitation as a function of frequency. The continuous line (blue) shows the scan of the bare resonator with a resonance frequency at 15.4~GHz, in good agreement with the simulations shown with dotted green line in the inset. When the resonator is loaded with a crystal, the resonance frequency shifts to $f_0=14.94$~GHz as shown by the scan at lower microwave power (red dots).}
	\label{fig:resonance}
\end{figure}

The switching probability is also highly sensitive to the presence of microwave radiation. When the SQUID is biased with a current pulse close to $I_{sw}$, microwave radiation can excite the SQUID and thus generate a switching event, similar to a thermal activation process \cite{Clarke2006,Lefevre-seguin1992}. One can therefore use the  micro-SQUID as both magnetic flux and microwave detector, to detect magnetization changes, the usual case of magnetometers, or microwave emission of spins, as in traditional ESR, respectively. If the magnetic flux change is larger than $\frac{1}{2}\Phi_0$, a feedback coil can be used to fix the SQUID's working point and track the change of $I_{sw}$\cite{Wernsdorfer1999,Chen2010}. Here, the flux changes due to spin excitations are expected to be very small and the working point location in the plot of Fig.~\ref{fig:mod} is essentially fixed by a normal static field arrising from an imperfect field alignement in the SQUID plan.

To verify the design, the SQUID $P_{sw}$ is measured at $T=$0.5~K under microwave excitation as a function of frequency, with and without a crystal of Gd-doped CaWO$_4$ placed on top of the chip. The CaWO$_4$ has a permitivitty\cite{Kim_JECS2006} $\epsilon_r\approx10$ and therefore a shift of the resonance frequency is to be expected when the crystal is placed on top of the chip. The blue line in Fig.~\ref{fig:resonance} shows the SQUID response in the case of an unloaded resonator with an estimated power at the chip entrance of $\approx -10$~dBm. Given the large amount of power, the bias current $I$ has a low value of 45~\textmugreek A, which allows to select the fundamental resonance at 15.4~GHz. The red dots show $P_{sw}$ for a loaded cavity and indicate a resonance shift to $f_0=14.94$~GHz. To show the versatility of the device, in the loaded case, the power is reduced by 35~dB, which allows to operate at a higher bias current (329~\textmugreek A). Outside the resonance, there are very few switching events ($P_{sw}\approx3\%$) likely due to thermal activation. Under these conditions, other very low intensity modes are observed above 15~GHz. One notes that the SQUID is not shunted by a designated capacitance, the modes are independent on bias $I$ and at a temperature of 0.5~K one cannot observe SQUID's internal level structure. During the ESR experiment, the microwave frequency is fixed at the main mode $f_0=14.934$~GHz and the external field can tune the spins in and out of resonance. The resonance is sufficiently broad to neglect in-plane field and microwave power induced shifts. 

To confirm the cavity effect we have performed finite element simulations in COMSOL, in which we discretized the structure of the chip numerically and computed the values of the S-parameters as a function of frequency. The SQUID was simulated as a COMSOL lumped port defined between the $\Omega$-loop and the ground metallic plane on which the Si chip is fixed. The port impedance defines the amount of absorbed energy and thus the resonance effect inside the waveguide structure.  For the physical dimensions of the current device, the optimal value is $Z_{eff}\approx142$~$\Omega$. The corresponding simulated frequency dependence of $S_{11}$ is given in Fig.~\ref{fig:resonance} (inset) and it shows a clear resonance at 15.4~GHz.  

\begin{figure}
	\centering
	\includegraphics[width=\mycol\columnwidth]{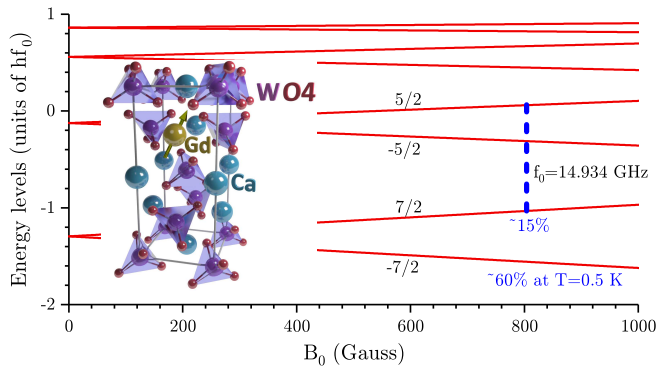}
	\caption{(color online) Gd$^{3+}$ energy levels for $B_0$ applied along the cristallographic $c$ axis, corresponding to the $z$-axis of Hamiltonian $H$. The blue dashed line shows the studied transition. At $T=0.5$~K and $B_0=804$~G, the first excited state $S_z=7/2$ has a thermal population of $\approx15\%$. The inset shows the tetragonal symmetry of the CaWO$_4$ crystallographic unit and a Gd$^{3+}$ spin (gold ball with an arrow). }
	\label{fig:level}
\end{figure}

\begin{figure}
	\centering
	\includegraphics[width=\mycol\columnwidth]{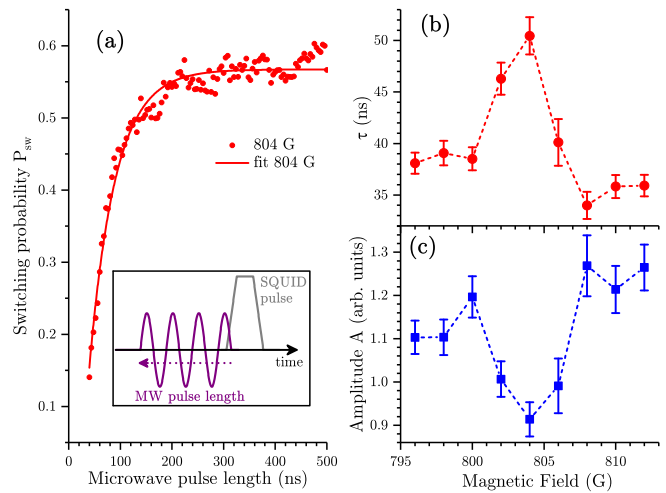}
	\caption{(color online) \textbf{(a)} $P_{sw}$ as a function of microwave pulse length at the resonance field 804~G. Continous lines are fits using exponential decays (see text). The inset is a sketch showing the timing of the microwave and the SQUID readout pulse (see text). Fit parameters $\tau$ \textbf{(b)} and A \textbf{(c)} as a function of $B_0$. At resonance field 804~G, $\tau$ is maximum, while $A$ has a minimum, indicating resonant absorption of energy by spins.}
	\label{fig:fitA}
\end{figure}


The spin Hamiltonian for the Gd$^{3+}$ $S=7/2$ ions diluted in CaWO$_4$ (tetragonal symmetry $I4/a$, see inset of Fig.~\ref{fig:level}) is the following\cite{Baibekov2017}: 
\begin{equation}
\label{eq:ham}
H=g_{\parallel}\mu_B S_z B_0+B_2^0 O_2^0+B_4^0 O_4^0+B_4^4 O_4^4+B_6^0 O_6^0+B_6^4 O_6^4,
\end{equation}
where $B_0$ is magnetic field along the crystallographic $c$-axis, $O_p^k$ are Stevens operators\cite{Stevens1952}, $\mu_B$ is Bohr magneton, $g_{\parallel}=1.991$ is the g-factor along the $c-$axis, and the crystal field parameters are (in units of MHz) $B_4^0=-1.14, B_4^4=-7.02, B_6^0=5.94\times 10^{-4}$ and $B_6^4=4.77\times 10^{-4}$. $B_2^0$ has a detectable temperature dependence\cite{HARVEY1971}. Instead of a value\cite{Baibekov2017} of $-894$~MHz at 293~K, our data indicates that at 0.5~K, its value should be $B_2^0=-897.8$~MHz. The obtained eigenvalues of $H$ are shown in Fig.~\ref{fig:level}; the blue dashed line indicates the field location of the resonance $S_z=7/2\leftrightarrow5/2$, at $B_0=804$~G. The shift in frequency caused by the loading of the resonator is larger than our expectations and no ground-state transitions fell within the appropiate experimental conditions. However, it is noteworthy that a successful measurement of such excited transition is relevant to show both the high sensitivity of the device as well as the potential of performing coherence control of multiple levels using this resonator and detection system.

At 6~K, the Gd spins have a typical relaxation time\cite{Baibekov2017} of the order of 10~ms and likely larger at our operating temperature of 0.5~K. Consequently the repetition frequency of the current pulse is decreased to 40~Hz to allow some amount of relaxation between consecutive measurements.  The microwave excitation can be applied in either CW or pulsed mode\cite{Chiorescu2003a}. The end point of the microwave pulse is chosen such that it barely overlaps with the micro-SQUID readout pulse, as sketched in the inset of Fig.~\ref{fig:fitA}(a). In this way, the SQUID can detect flux changes due to spin rotation as well as energy absorption without saturating the SQUID readout. Once the overlap is fixed, the length of the microwave pulse can be varied by changing its start time, as indicated with a dashed left-arrow. For each length, the value of $P_{sw}$ is obtained  by repeating the shown pulses for 3,000 times and recording the percentage of SQUID switching events. The current SQUID pulse is fixed at $\approx303$~\textmugreek A and 2.1~\textmugreek s.

A single crystal of Gd-doped CaWO$_4$ with dimensions of 0.63 mm $\times$ 2.4 mm $\times$ 0.95 mm was placed over both shortcircuits shown in Fig.~\ref{fig:device} such that the external magnetic field is parallel to its $c$-axis (also $z$-axis) with a precision of $\approx2^\circ$. Using dedicated direct-current and high frequency electronics, the SQUID switching probability can be recorded as a function of temperature, magnetic field and microwave pulse characteristics. At $T=0.5$~K, frequency $f_0$ and power $\approx-10$~dBm, Fig. \ref{fig:fitA}(a) shows $P_{sw}$ vs. microwave pulse length $t_{mw}$ at the resonance field $B_0=804$~G. As $t_{mw}$ increases, the amount of energy pumped in the SQUID increases and therefore $P_{sw}$ increases up to a saturation level reached around $t_{mw}\approx300$~ns. The $P_{sw}(t_{mw})$ data is fitted with function $P_{sw}=C_0-A\exp^{-t_{mw}/\tau}$ where $C_0, A$ and $\tau$ are fit parameters, as shown in Fig.~\ref{fig:fitA}(a) with a continous line. The measurements are repeated with a 2~G step in the resonance region and the obtained $A$ and $\tau$ parameters and their uncertainties are shown in Fig.~\ref{fig:fitA}(b) and (c). 

At the resonance field 804~G, the characteristic time $\tau$ has a peak, while $A$ has a dip, which are due to the absorption of energy by resonant spins (slower saturation towards a smaller plateau). This indicates the resonant absorption of microwave energy corresponding to the $S_z=7/2\leftrightarrow5/2$ transition. The linewidth of the resonance signal is about $\sim$4~G or 25~MHz in frequency units, which is a typical value for diluted spin systems (decoherence time of the entire spin ensemble $T_2^\star\approx$40 ns). Spin-echo detection and Rabi oscillations at higher microwave powers are needed to characterize the decoherence time. The SQUID-waveguide device discussed here does offer the flexibility needed for such future studies.   
   
The microwave pulse can excite only the spins situated in the very close proximity of the chip surface. The numerical simulations show that the microwave field inside the $\Omega$-loop decays fast with height: from a surface value of 0.15~G (corresponding to an estimated -10~dBm input power in our experiments), the field halves at $\sim$11~\textmugreek m above the chip. Due to the $\sim$\textmugreek m size of the SQUID and the fast decay of the microwave field, the effective detected sample volume is estimated as in Ref.~[\onlinecite{Toida2016}] to be $\sim$\textmugreek m$^3$ size, which corresponds to $\sim10^7$ spins in the case of the CaWO$_4$:Gd$^{3+}$ sample discussed here.

To perform Rabi oscillations with well-defined nutation rates, it is important to have an homogenous microwave field in the sample. This can be achieved by using very thin samples, which can be prepared, for instance, by chemical growth in 2D configurations, using radiation to generate defects or by cuting ultra-thin samples using a focused ion beam. Typical decoherence times $T_2$ in diluted spin systems are of the order of $\sim1-10$~\textmugreek s\cite{Baibekov2017,Bertaina2009}, requiring Rabi frequencies of $\sim10$~MHz and microwave fields of several Gauss. Such values are attainable in future implementations of the technique presented here, by using higher microwave pulsed fields while operating the  SQUID in its classical regime ($T\gtrsim0.1-1$~K). It is expected that the signal size will actually increase due to population inversion generated by full spin rotation.  

We present a sensitive on-chip detection scheme able to measure the first excited transition in a multi-level quantum spin system, at low temperatures. Gd spins $S=7/2$ diluted in a CaWO$_4$ crystal are in resonance with the microwave excitation contained in a small volume around an $\Omega$-shaped shortcircuit. A rough estimation of the number of detected spins is $\sim10^7$. This SQUID-based resonator can be used in future studies to perform gated control of spin rotations.

This work was supported by the NSF Grant No. DMR-1206267 and CNRS-PICS CoDyLow. The NHMFL is
supported by the Cooperative Agreement Grant No. DMR-1157490 and the State of Florida. We acknowledge support from the Strategic Priority Research Program of the Chinese Academy of Sciences (Grant No. XDB04000000). We are thankful to Dr. A. Tkachuk for providing the Gd sample and to X. Lian, S. Zhang and Dr. Xiong for the support in device fabrication.

\bibliography{ref}

\end{document}